# Electron Groups in Solar Wind and Gas Discharge Plasmas

**Vladimir Kolobov,** [a,b,1] **Robert Arslanbekov** [a] **and Dmitry Levko** [a]

[a] CFD Research Corporation, Huntsville, AL 35806, USA
[b] University of Alabama in Huntsville, AL 35899, USA

E-mail: vladimir.kolobov@cfdrc.com

**Abstract**. The formation of electron groups is a common phenomenon in gas discharges and space plasmas. We describe similarities between electron kinetics in solar wind, cathode region of glow discharges, and magnetic nozzles. Remarkable resemblances between adiabatic focusing of magnetized electrons and spherical geometry effects for neutral particles escaping planet's atmosphere are pointed out. The previously developed kinetic solvers for electrons coupled to Poisson equation for self-consistent simulations of the electric field are used to analyse the formation of electron groups in the cathode region of glow discharges and solar wind plasma.

## 1. Introduction

Electron velocity distribution function (VDF) in plasma is rarely a Maxwellian.[1,2] There are several reasons for non-Maxwellian VDFs for both fully-ionized space plasmas and weakly-ionized gas-discharge plasmas. In the first case, magnetized electrons, often partially confined by plasma-produced electric fields, are subject to wave-particle interactions and turbulence, which dominate over Coulomb interactions among charged particles. In the second case, electron heating by external electric fields and collisions with neutral plasma species produce peculiar non-equilibrium conditions for most low-temperature bounded plasmas. In this paper, we discuss typical scenarios for the formation of weakly-coupled electron groups in plasmas and show examples of kinetic simulations of electrons.

The main reason for non-equilibrium distribution of electrons in plasmas is their tiny mass with respect to ions and neutrals. As a result, electrons move faster than heavy particles, even for the same mean energy (temperature) of both species. In spatially non-uniform plasma, this creates an electric field to maintain quasi-neutrality. The field decelerates electrons and accelerates ions to equalize their fluxes. In bounded plasmas, one group of the electrons is got trapped by the electric field in the plasma volume, while the other group can escape the potential well and reach the plasma boundaries. The trapped and free electrons form the two quite independent groups. Between plasmas with different properties, an electric double layer is often formed to equalize the fluxes of electrons and ions at interfaces between these plasmas. The double layers formed in both collisionless and collisional plasmas, create groups of electrons and ions with different properties.

The second reason for the formation of electron groups in collisional plasmas are the peculiar properties of electron collisions with neutrals and Coulomb interactions among charged species. Due to the mass discrepancy between the colliding partners, elastic collisions of electrons with neutrals result in large momentum change of the electrons and only in small changes of their energy. Slow electrons with energies lower than the inelastic energy thresholds of atoms and molecules can be easily accelerated by the electric field, but their energy exchange with heavy species is greatly suppressed. As a result, electron temperature in gas-discharge plasma typically exceeds the temperature of heavy species by two orders of magnitude. On the other hand, fast electrons with energies exceeding the electronic excitation threshold of atoms and molecules (about 10 eV) lose large portion of their energy. With further increase of the electron kinetic energy, electron scattering becomes highly anisotropic, and the energy loss in collisions passes through a maximum with respect to the electron energy. High energy (runaway) electrons can be continuously accelerated by the electric field forming highly anisotropic electron beam. Thus, the fast and slow electrons form two groups of electrons with different kinetics.

---

[1] To whom any correspondence should be addressed.

In this paper, we describe some typical scenarios for the formation of electron groups in solar wind and laboratory plasmas. The previously developed Fokker-Planck kinetic solvers [3,4] are used to illustrate peculiarities of electron kinetics in the cathode region of glow discharges and solar wind.

## 2. Electron Kinetics in Collisional Plasma of Gas Discharges

Collisional plasma of gas discharges is created by the electron-impact ionization of neutral gas background. Being accelerated by the externally applied electric fields, electrons scatter in elastic collisions with gas molecules converting their directed kinetic energy into thermal energy of chaotic motion, aka electron heating. The mean electron energy (temperature) in low-temperature plasma (LTP) can exceed the gas temperature by two orders of magnitude. The electron drift velocity remains much smaller than the thermal velocity (i.e. near-isotropic VDF). Electrons are responsible for complex energy flow in LTP by converting their thermal energy into excitation of rotational and vibrational states of molecules, dissociation of molecules, excitation of electronic states of atoms and molecules, ionization processes, and electron-induced chemical reactions. Furthermore, they are responsible for creating electric fields in plasma, which maintain plasma quasi-neutrality.

Cathode region of DC glow discharges is an excellent example of electron self-organization into groups in collisional plasma.[5] The main function of the cathode region is to generate the flux of electrons by multiplication of the primary electrons emitted by the cathode. Substantial ionization is required near the cathode to accelerate the primary electrons within the cathode sheath and produce electron avalanche. From visual observation, several cathode regions are distinguished, which for historical reasons are called negative glow, Faraday dark space, etc.

Figure 1 illustrates schematically the formation of three electron groups in the cathode region. The first group includes fast electrons accelerated in the cathode sheath and responsible for enhanced ionization in the negative glow, at $x < \Lambda$, where $\Lambda$ is the fast electron penetration range. This group includes the primary electrons injected from the cathode and secondary electrons produced in the cathode sheath, at $x < d$, due to the gas ionization. Fast electrons are characterized by strongly anisotropic VDF.

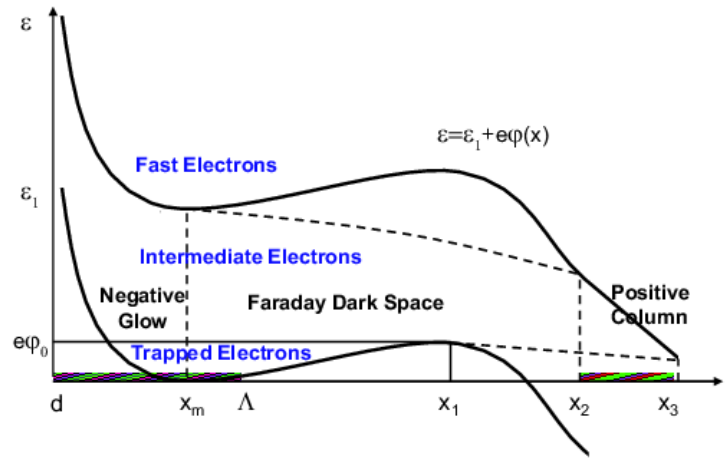

*Figure 1. Potential profile in the cathode region and three groups of electrons.*

The second group includes electrons produced in the negative glow outside the sheath, at $x > d$. These "intermediate electrons" are responsible for the current flow in the Faraday Dark Space (FDS) between the points $x_1$ and $x_2$. The intermediate electrons have near-isotropic VDF. They are also responsible for the ionization processes in the region $x > x_2$, and the FDS transition into positive column.

The third group of electrons includes the electrons trapped by the electrostatic potential near the point $x_m$ where the electric potential has a local minimum. The trapped electrons do not contribute to the current flow, but they are responsible for a sharp peak of plasma density in the cathode region. The VDF of trapped electrons is often Maxwellian, their density is defined by the Boltzmann relation, and the temperature is slightly above the gas temperature.

We have applied the previously developed 1d1v and 1d2v kinetic solvers [3,4] to simulate the formation of electron groups in the cathode region. The kinetic equations were solved using Adaptive Mesh in Phase Space (AMPS) technique with spherical coordinate system in velocity space. Elastic collisions of electrons with neutral atoms, excitation of atomic levels and electron-impact ionization were taken into account. We solved kinetic equations using Finite Volume method with adaptive Cartesian mesh without the splitting physical and velocity spaces. For transport in configuration (physical) space, we could use Cartesian, cylindrical or spherical coordinate systems depending on the problem type. Kinetic solvers for electrons were coupled to the Poisson equation for the electric field.

For coupling kinetic and Poisson solvers, which operate in spaces with different dimensions, the quad/octree (1d1v/1d2v) phase-space grid is projected into 1d Poisson-solver grid using cell-centred x-locations. The calculated particle densities are then transferred from the phase-space grid to the 1d space grid. After solving the Poisson equation on the 1d grid, electric potential and electric field are returned back at cell centres and faces of the phase-space grid.

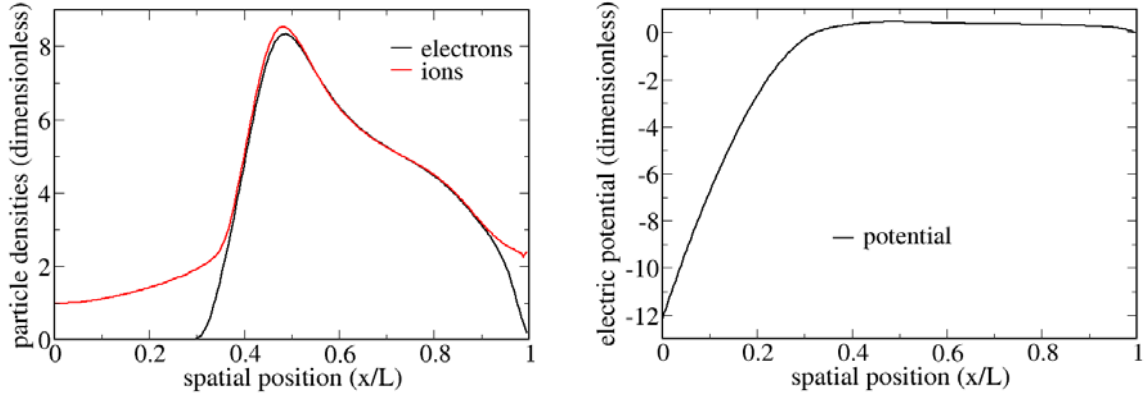

*Figure 2. Electron and ion densities (left) and the electrostatic potential for a short glow discharge.*

Figure 2 shows results of self-consistent hybrid simulations of a short glow discharge using the 1d1v Fokker-Planck solver for electrons, fluid model for ions, and Poisson equation for the electrostatic potential. The formation of space charge sheath near the cathode, quasineutral plasma in the middle of the gap and a space charge near the anode is clearly visible. The formation of the potential well for electrons, the peak of plasma density near the potential maximum, and gradual change of the density profile can also be seen.

Simulations shown in Figure 2 were obtained using the explicit code and helped us to better understand the challenges associated with disparate time scales for electrons and ions. The time step in this code was limited by the CFL criterium of the electron kinetic solver. The electron VDF and the potential well trapping electrons were formed on the (fast) electron time scale. At the same time, the spatial distribution of the plasma density evolved on the (slow) ion time scale (via ambipolar diffusion).

As a result, using explicit hybrid plasma solver required about $10^8$ time steps for these simulations. This indicated that explicit plasma solvers are not practical for solving even one-dimensional plasma problems evolving on the ion time scale. Our current efforts are devoted to the implementation of implicit hybrid solvers capable of simulating collisional plasma processes occurring on the ion time scales.

### 3. Electron Kinetics in Solar Wind

Measured electron VDFs in solar wind contain three groups of electrons: core, halo and strahl. The Maxwellian "core" comprises the bulk of the electron density. The strahl is a field-aligned beam with energies from 10 eV and 1 keV at 1 AU. The halo is a nearly-isotropic component in the same energy range. It is commonly assumed that the strahl and halo formed as a result of adiabatic electron focusing in spatially weakening magnetic field and its broadening via pitch-angle scattering. Details of the pitch-angle scattering and relative contributions of different scattering mechanisms are still under debates. The low energy scattering can be provided by Coulomb collisions, the high energy scattering could be due to the wave-particle interactions (such as whistler-mode turbulence). Self-consistent kinetic simulations of VDFs taking into account all the relevant processes are still missing.

*3.1. Kinetic equation for electrons in magnetized plasmas*

The Fokker-Planck kinetic equation for the gyro-averaged velocity distribution function of electrons has the form:[6]

$$\frac{\partial f}{\partial t} + v\mu\frac{\partial f}{\partial s} + \frac{v}{2L(s)}(1-\mu^2)\frac{\partial f}{\partial \mu} - \frac{eE}{m}\left[\frac{1-\mu^2}{v}\frac{\partial f}{\partial \mu} + \mu\frac{\partial f}{\partial v}\right] = \nu_{ee}\frac{\partial}{\partial \mu}(1-\mu^2)\frac{\partial f}{\partial \mu} \qquad (1)$$

where $s$ is the distance along the magnetic field line, $\mu = v_s/v$ is the pitch angle cosine, $L(s) = -(d\ln B/ds)^{-1}$ is the characteristic length of adiabatic focusing by the magnetic field, $E$ is the electric field component along the magnetic field line, and $\nu_{ee} \sim n(r)/v^3$ is the Coulomb collision frequency. For the Parker spiral magnetic field model, the adiabatic focusing length is:[7]

$$L(s) = r\frac{\left(1+\left(\frac{\Omega r}{V}\right)^2 \sin^2\theta\right)^{3/2}}{2+\left(\frac{\Omega r}{V}\right)^2 \sin^2\theta} \qquad (2)$$

where $V$ is the solar wind speed, $\Omega$ is the annular rotation velocity of the Sun, and $r$ and $\theta$ are the coordinates of the heliocentric spherical coordinate system. For $\theta = 0$ or for $\Omega r/V \ll 1$, one obtains $L(s) = r/2$, which corresponds to the magnetic monopole with radial magnetic field ($s = r$, and $\frac{d\ln B}{dr} = -2/r$). For a magnetic dipole, $\frac{d\ln B}{ds} = -3/s$.

It is remarkable that the left side of the kinetic equation (1) for magnetized electrons resembles the collisionless kinetic equation for neutral particles in gravitational field:[8]

$$\frac{\partial f}{\partial t} + v\mu\frac{\partial f}{\partial r} - \frac{v}{r}(1-\mu^2)\frac{\partial f}{\partial \mu} - g\left[\frac{1-\mu^2}{v}\frac{\partial f}{\partial \mu} + \mu\frac{\partial f}{\partial v}\right] = 0 \qquad (3)$$

where $g = -\gamma m/r^2$, and $\gamma$ is the gravitational constant. Lie-Svendsen [9] has pointed out that the effect of spherical geometry for neutral particles results in a force that has exactly the same form as the adiabatic focusing force for a magnetic monopole and is 2/3 of the magnetic dipole force. Indeed, for charged particles, the electric field $E(r)$, which is formed in plasma to balance fluxes of electrons and ions, must be calculated self-consistently with the particle kinetics.

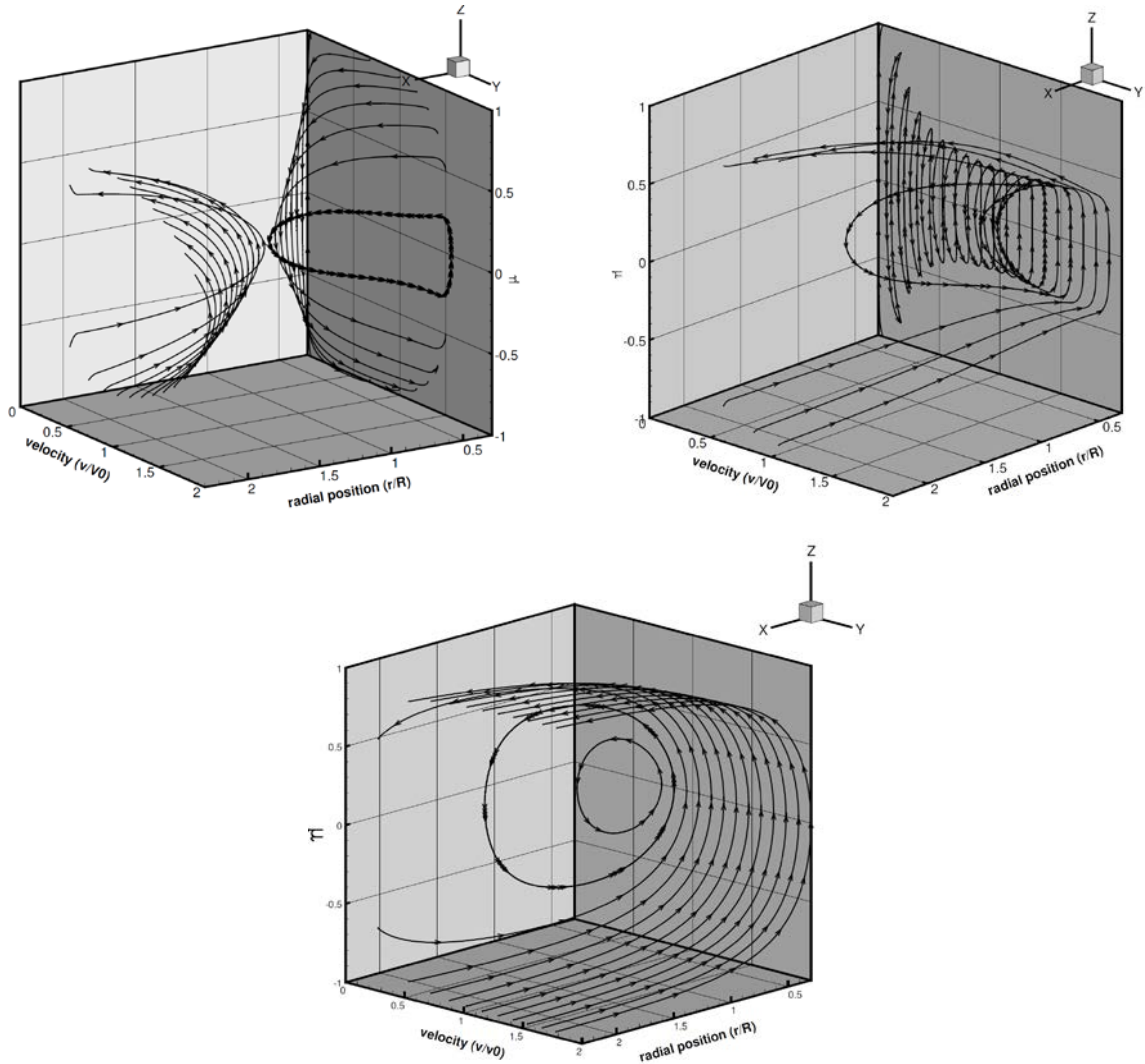

*Figure 3. Characteristics of the neutral particle motion in gravitational field of a planet (top) and magnetized solar wind electrons trapped by an electrostatic field (bottom).*

Figure 3 shows calculated characteristics of neutral particles in a gravitational field of a planet[10] and characteristics of adiabatically focused magnetized electrons in a prescribed electric field. It is seen that in both cases three groups of the particles are present: free particles, which are evaporated from the boundary and escaping the gravitational or electrostatic forces; reflected particles, evaporated from the boundary and returned back by these forces; and trapped particles, which are bouncing within the potential well formed by the retarding forces from one side and the mirror effect in the vicinity of the boundary. It is important to point out the significant difference between the motions of neutral particles and magnetized electrons. In the first case, closed trajectories correspond to the particle motion on the elliptic trajectories around the center of gravity. In the second case, the electrons move along the magnetic field lines (in our case, in the radial direction) turning back when their radial velocity becomes zero. In both cases, the populations of free and reflected electrons are totally determined within the collisionless model, but the population of trapped particles can be found only by taking collisions into consideration.

### 3.2. The Strahl

An analytical solution of the kinetic equation (3) for spherical geometry is well-known in the molecular gas dynamics:[8]

$$f(r,v,\mu) = \begin{cases} f_0(r_0, v, \mu), & \sqrt{1-\left(\frac{r_0}{r}\right)^2} < \mu < 1 \\ 0 & -1 < \mu < \sqrt{1-\left(\frac{r_0}{r}\right)^2} \end{cases} \quad (4)$$

The VDFs calculated according to Eq. (4) are shown in Figure 4. The VDF becomes highly anisotropic with decreasing $r_0/r$ ratio. At $r/r_0 \sim 200$, which correspond to 1 AU from the Sun, the VDF is concentrated at $\Delta\mu \leq 0.5(r_0/R)^2 \sim 10^{-5}$. Resolving such an VDF with Cartesian mesh in $(v, \mu)$ velocity space requires mesh with a cell size $1/2^n \; 10^{-5}$, which requires the refinement level $n = 16$-$17$.

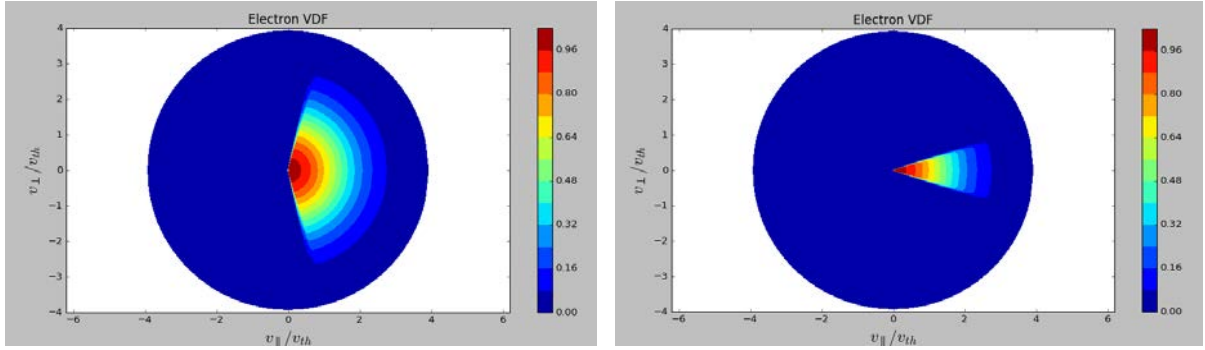

*Figure 4. The VDFs calculated according to Eq. (4) at two locations: $r_0/r \sim 0.95$ and $r_0/r \sim 0.25$*

### 3.3. The Halo

Pitch-angle scattering of solar wind electrons results in the broadening of the strahl and the halo formation. Effects of Coulomb collisions on the halo formation were analyzed in Horaites et al.[11], who has shown that the main properties of the strahl are well correlated with the Coulomb collision model. However, the width of the strahl is typically larger than that predicted by the Coulomb collisions for high energies.

In our simulations reported below, the boundary condition for VDF at $r = r_0$ was defined as

$$f(r_0, v, \mu) = exp(-(v/v_0)^2), \quad \mu > 0 \quad (5)$$

There is no need for the boundary condition for $\mu < 0$ at $r = r_0$, as long as there is no diffusion along the $r$-direction in Eq. (1). The boundary condition at $r = L$ corresponds to the absence of electron flux to the Sun. In our simulations below, $v_0 = 0.5$, and $v_{ee} = C/(r^2 v^3)$. Results of simulations for $C = 0.025$ and $r_0 \sim 0.1$ are shown in Figures 5 and 6. In Figure 5, the electric field is small, $E_0 \sim 0.01$, and the trapping effect of the electric field is negligible. In Figure 6, for $E_0 \sim 0.1$, the trapping effect of the electric field becomes substantial.

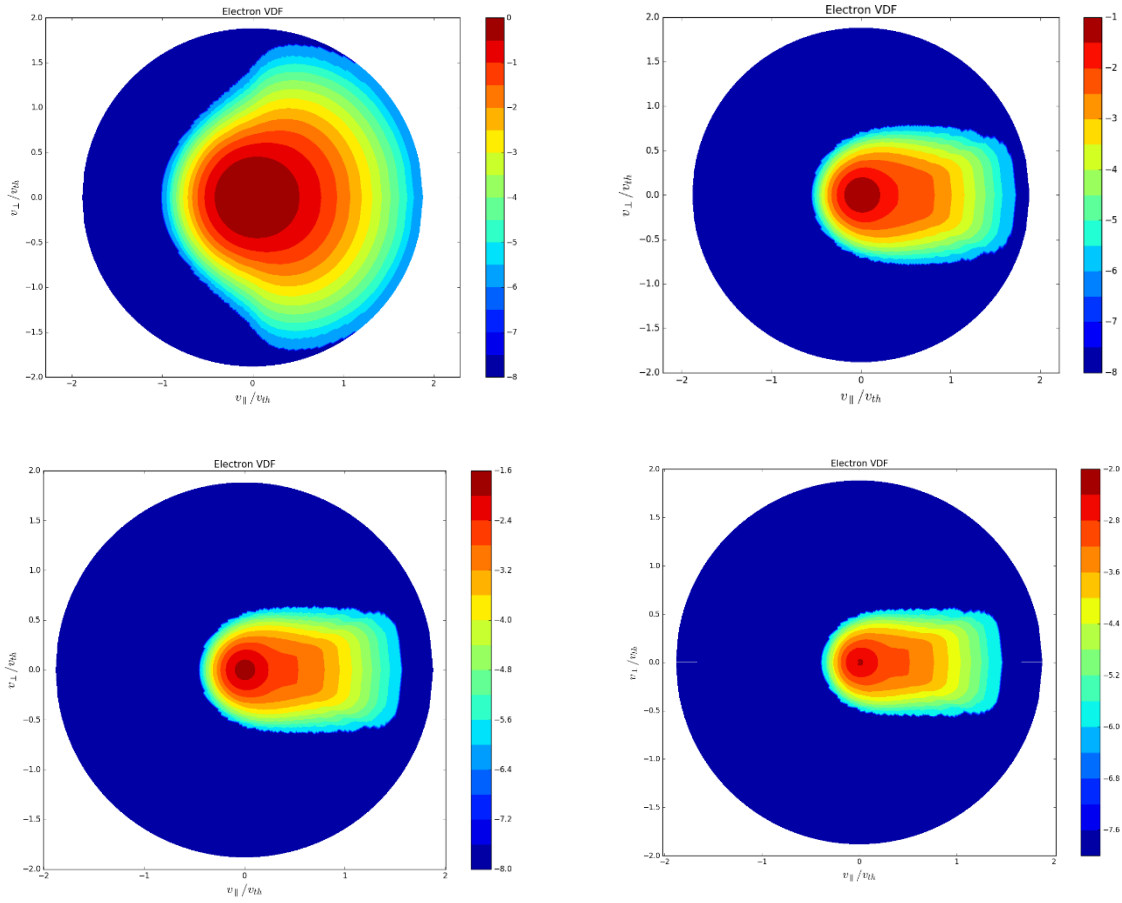

*Figure 5. VDF at $x = 0.02, 0.5, 1.0, 1.5$ for C=0.025 for $E_0 \sim 0.01$.*

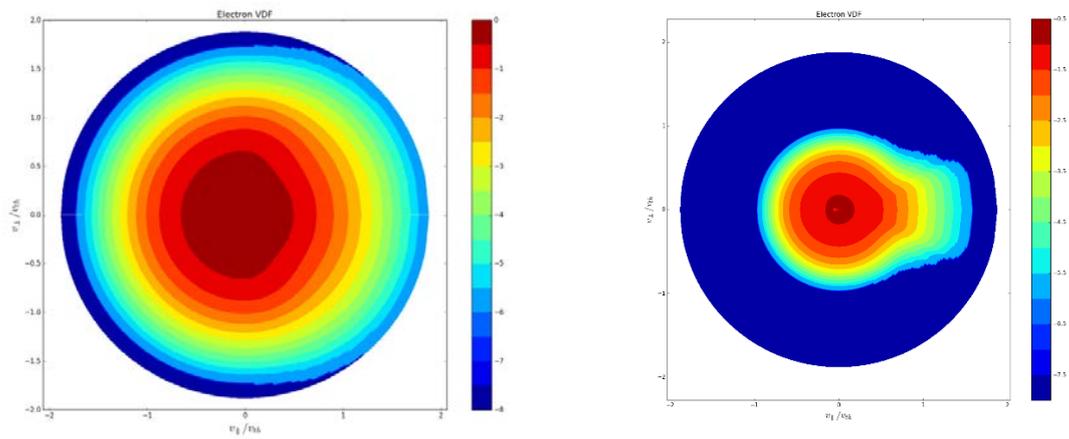

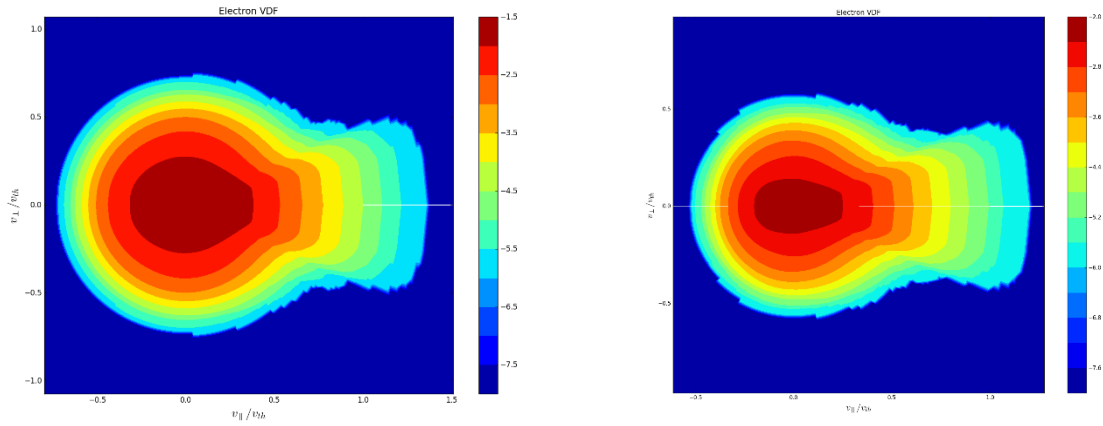

*Figure 6. VDF at $x = 0.02, 0.5, 1.0, 1.5$ for C=0.025 for $E_0 \sim 0.1$.*

Figure 7 shows the calculated spatial distributions of electron density, mean velocity and temperatures for different values of the electric field. It is seen that the electron flux changes its sign for the large values of the electric fields. This means that the electric field we assumed is too strong at low distances.

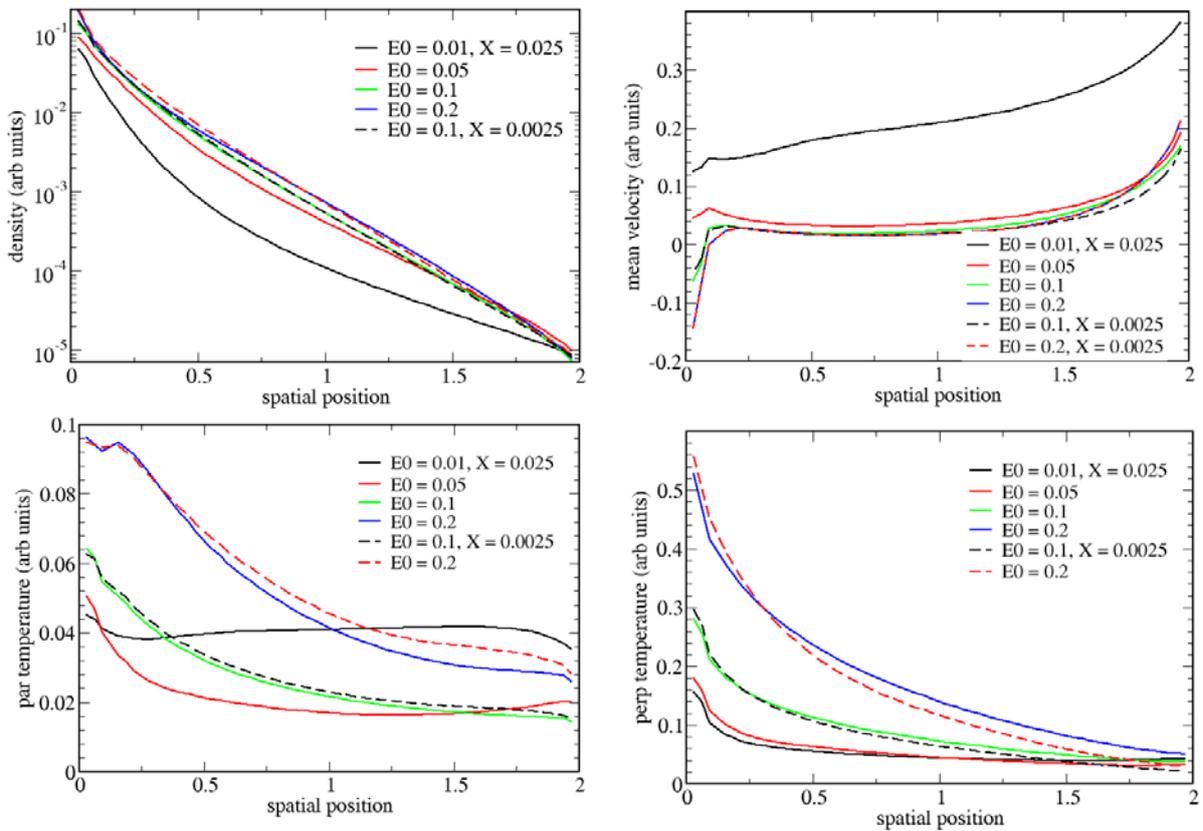

*Figure 7. Density, mean velocity and temperatures for different values of the electric field and plasma density.*

We have also performed self-consistent calculation of the electric field assuming zero density of ions for the same conditions shown above. The injected flux of electrons was varied to analyse the electrostatic trapping of the electrons. Figure shows the electron characteristics in self-consistent electric field. It is seen that the three groups of electrons are formed as in the previous cases: fast electrons adiabatically focused along the field ($\mu = 1$), the electrons returning back to the injection surface (see closed lines of the left figure), and trapped electrons, which have closed characteristics in the right figure). All characteristics in both figures cross the $\mu = 0$ line. The left figure corresponds to the characteristics crossing the $r = r_0$ plane.

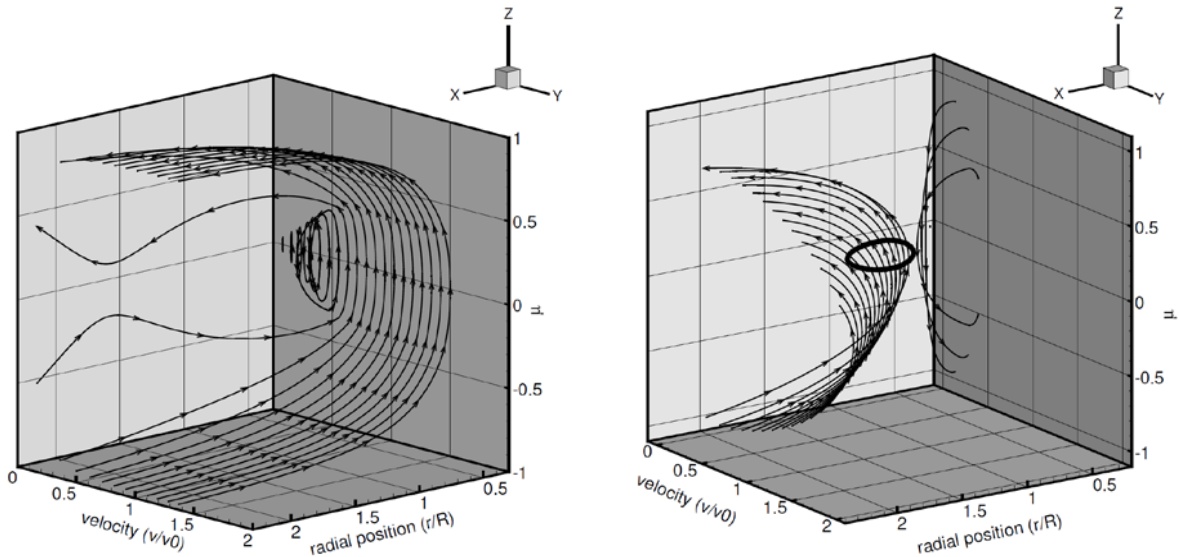

Figure 8. Characteristics of the collisionless electron motion in self-consistent electric field

Figure 9 shows the radial distribution of electron density and electric potential for different values of the injected flux. It is seen that with the increasing flux the value of the trapping field increases and the larger number of electrons become trapped in the potential well formed by the electric potential. It should be reminded that Coulomb scattering has been included in these calculations otherwise the characteristics corresponding to trapped particles were empty.

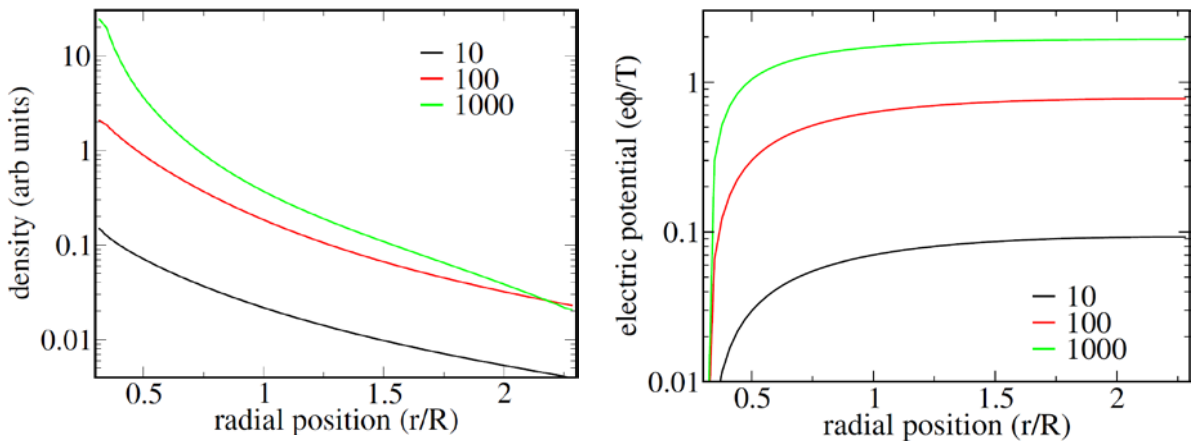

Figure 9. Electron density (left) and electric potential (right) for different values of the injected flux.

## 4. Electron Groups in Magnetic Nozzles

Magnetic nozzles contain a divergent magnetic field capable of guiding and accelerating a magnetized plasma jet into vacuum. To be usable for electric propulsion, the plasma jet must detach from the applied magnetic field downstream to form a free expanding plasma plume. Understanding the plasma detachment process has remained elusive for decades. In a diverging magnetic field, the electron thermal energy is transformed into directed energy of ions via an ambipolar electric field. A potential well for electrons is formed inside the nozzle by a combination of external magnetic field and self-generated ambipolar electrostatic potential. The axial forward-motion of electrons is governed by two forces: the electric one, which decelerates them, and the adiabatic magnetic focusing mirror effect. Three groups of electrons have been found in a divergent nozzle: free electrons, emitted by the plasma source and lost downstream (which constitute the electron current); reflected electrons, emitted by the source and returned back to the source; and trapped electrons, bouncing within the potential well. While the populations of free and reflected electrons are totally determined within collisionless model in terms of B(s) and f(s), the analysis of trapped electrons requires taking collisions into consideration. Ions expanding supersonically can be described using a simple cold fluid model. For electrons, an 'equivalent' local polytropic coefficient defined as $\gamma = d \ln p_e / d \ln n_e$ (where $p_e$ is the electron pressure) shows large variations along the plume. Near the source electrons are near-isothermal, then tend to an adiabatic behavior, and finally to a (cold) isothermal behavior again.[12]

Electric double layers with non-monotonic distributions of the electrostatic potential have been often observed in expanding plasmas. Current-free double layers have been found in low-pressure expanding magnetized plasmas of magnetic nozzles. Although double layers have been mostly studied for collisionless plasmas, they also appear in collisional plasma (both current-carrying and current-free). The cathode region of DC discharges with two field reversals in the negative glow and Faraday dark space described above and the anode region of DC discharges are the typical examples. The collisional double layers have been found in hollow cathode discharges and plasma thruster nozzles.[13] Collisional double layers have been also observed in nonlinear striations (ionization waves) in DC glow discharges. In all these cases, the presence of double layers indicates the appearance of different electron groups, which should be taken into consideration for the proper problem analysis.

## 5. Conclusions

We have illustrated that the formation of electron groups due to trapping of electrons by electric fields is a common phenomenon in gas discharges and space plasmas. We have also illustrated remarkable resemblances between adiabatic focusing of magnetized electrons and spherical geometry effects for neutral particles escaping planet's atmosphere. The previously developed 1d1v and 1d2v kinetic solvers have been applied to simulate the formation of electron groups in the cathode region of glow discharges and solar wind plasma. Kinetic solvers for electrons were coupled to Poisson equation for self-consistent simulations of the electric field. Computational challenges associated with disparate time scales for electrons and ions and the need for implicit plasma solvers have been discussed.

In our simulations, Coulomb scattering has been included but the energy drift and diffusion responsible for the Maxwellization of low-energy electrons has not been included yet. Including full effects of Coulomb interactions for the calculation of the VDF and self-consistent simulations of the electric field and ion space charge is planned for our further work. We also plan to add cross-field diffusion terms into Fokker Planck kinetic equation for magnetized electrons to describe electron demagnetization and plasma detachment in magnetic nozzles.

**Acknowledgments**

This work was partially supported by the DOE SBIR Project DE-SC0015746, by the US Department of Energy Office of Fusion Energy Science Contract DE-SC0001939, and by the NSF EPSCoR project OIA-1655280 "Connecting the Plasma Universe to Plasma Technology in Alabama: The Science and Technology of Low-Temperature Plasma".